\documentclass
[floatfix,superscriptaddress,secnumarabic,amssymb,amsmath,nobibnotes,aps,prd,showkeys,nofootinbib,onecolumn,notitlepage,10pt]{revtex4}%
\usepackage{setspace}
\usepackage{xcolor}
\usepackage{amsmath}
\usepackage{amsfonts}
\usepackage{amssymb}
\usepackage{verbatim}
\usepackage{graphicx,bm}
\usepackage[caption=false]{subfig}
\usepackage[colorlinks]{hyperref}
\usepackage{graphicx}%
\providecommand{\U}[1]{\protect\rule{.1in}{.1in}}

\newcommand{\be}{\begin{equation}}
\newcommand{\ee}{\end{equation}}

\newcommand{\mincir}{\raise
-3.truept\hbox{\rlap{\hbox{$\sim$}}\raise4.truept\hbox{$<$}\ }}
\newcommand{\magcir}{\raise
-3.truept\hbox{\rlap{\hbox{$\sim$}}\raise4.truept\hbox{$>$}\ }}

\hypersetup{
breaklinks=true,
pdfstartview={FitH},      colorlinks=true,           linkcolor=blue,              citecolor=red,            filecolor=magenta,          urlcolor=blue,               anchorcolor=green,          linktocpage=true
}
\ifx\pdfoutput\relax\let\pdfoutput=\undefined\fi
\newcount\msipdfoutput
\ifx\pdfoutput\undefined\else
\ifcase\pdfoutput\else
\ifx\paperwidth\undefined\else
\ifdim\paperheight=0pt\relax\else\pdfpageheight\paperheight\fi
\ifdim\paperwidth=0pt\relax\else\pdfpagewidth\paperwidth\fi
\fi\fi\fi
\begin{document}
\title{Unifying the Dark Sector with the New Generalized Chaplygin Gas: Observational Constraints}
\author{Abdulla Al Mamon}
\email{abdulla.physics@gmail.com}
\affiliation{Department of Physics, Vivekananda Satavarshiki Mahavidyalaya (affiliated to
the Vidyasagar University), Manikpara-721513, West Bengal, India}
\author{Andronikos Paliathanasis}
\email{anpaliat@phys.uoa.gr}
\affiliation{Institute of Systems Science, Durban University of Technology, Durban 4000,
South Africa}
\affiliation{Departamento de Ciencias Matem\'{a}ticas y F\'{\i}sicas, Universidad Catolica
de Temuco, Temuco, Chile}
\affiliation{National Institute for Theoretical and Computational Sciences (NITheCS), South Africa}
\author{Subhajit Saha}
\email{subhajit1729@gmail.com}
\affiliation{Department of Mathematics, Panihati Mahavidyalaya, Kolkata 700110, West
Bengal, India}

\begin{abstract}
In light of recent cosmological observations, we examine a generalized
Chaplygin gas model with a redshift-dependent exponent as a framework for
describing the dark energy and dark matter content of the Universe.
Specifically, we treat this fluid as a single unified component and test it
against late-time background observational data. We employ Type Ia supernova
data, cosmic chronometers, and baryon acoustic oscillations from the second
data release of the Dark Energy Spectroscopic Instrument. We perform a
Bayesian analysis for parameter estimation and compare the model with
$\Lambda$CDM. We find that the generalized Chaplygin gas provides
systematically higher values of the combined likelihood; nevertheless, once
the larger number of free parameters is taken into account, both the Bayesian
evidence and the Akaike Information Criterion suggest that the model is
statistically indistinguishable from $\Lambda$CDM.

\end{abstract}
\maketitle




\section{Introduction}

\label{sec1}
Recent cosmological observations indicate that the Universe is currently
undergoing a phase of rapid acceleration
\cite{acc1,acc2,acc3,acc4,acc5,p2015,p2018}. In the context of General
Relativity, this accelerated expansion is widely believed to be driven by the
so-called dark energy (DE), which dominates the present cosmic energy budget.
The simplest and most commonly adopted model for DE is the $\Lambda$CDM model,
in which the cosmological constant $\Lambda$ couples only gravitationally to
cold dark matter (DM). The latter is assumed to behave as pressure-free dust;
its nature remains unknown, and its presence is inferred exclusively through
gravitational effects. For a recent discussion, we refer the reader to Ref.
\cite{orl1}.\newline

The physical origin and nature of these two fluids, that is, of DE and DM, are
still unknown. The cosmological constant $\Lambda$, describes the vacuum
energy density, and possesses an effective fluid description where the
pressure component is analogous to the energy density. The proportionality
constant is known as the equation of state parameter (EoS), and takes the
constant value $w_{\Lambda}=-1$. Although this model agrees well with a wide
array of observational data, it is affected by several well-known theoretical
issues such as the fine-tuning problem and the cosmic coincidence problem
\cite{ftplcdm,ccplcdm}. On the other hand, recent high-precision cosmological
observations have revealed notable discrepancies, particularly the $H_{0}$ and
$S_{8}$ tensions, which indicate possible limitations of the $\Lambda$CDM
framework. The $H_{0}$ tension refers to the difference between the Hubble
constant derived from Planck CMB data within the $\Lambda$CDM framework
\cite{p2018} and a higher value measured from local distance-ladder
observations by the SH0ES network \cite{refh02}. Additionally, recent
observations from the Dark Energy Spectroscopic Instrument (DESI)
\cite{DESI:2025zpo,DESI:2025zgx} suggest a dynamical nature of the DE EoS,
introducing new challenges for the $\Lambda$CDM framework.\newline

These challenges associated with the cosmological constant $\Lambda$, have motivated the
development of several alternative DE models, in a phenomenological and
theoretical perspective, including the introduction of quintessence fields,
k-essence, phantom, and tachyon field models among others \cite{de1,de2}.
Nevertheless, understanding the fundamental nature and the origin of DE
remains one of the major unresolved problems in Cosmology. For a detailed
overview of the $\Lambda$CDM model and its problems, see for instance Ref. \cite{revlcdm}.\newline

The shortcomings of the $\Lambda$CDM framework have encouraged the search for
alternative cosmological models that are simpler and more unified. A natural
line of inquiry is whether a single dark fluid can simultaneously describe
both DM and DE. Unified models of this type offer distinct advantages. For
one, they require only one component to account for both the late-time cosmic
acceleration and the development of cosmic structures. Additionally, they
allow a consistent treatment of DM and DE at the perturbation
level~\cite{Hu1}. A single dark fluid that accounts for both DM and DE is
generally referred to as a unified dark fluid. Such fluids have become a major
topic of interest in contemporary cosmology due to their ability to describe
both DM and DE under a unified theoretical description. Detailed discussions
and reviews on this subject can be found in Refs. \cite{refum1,refum2,refum3,gcg1,gcg2,ngcg,Sen1,refmcg1,refmcg2,refmcg3,refmcg4,refmcg5,sahampla,ldf1,ldf2,ldf3,ldf4,ldf5,ldf6,lw1,lw2,lw3,lw4,uni1,uni2,uni3,uni4,uni5,uni6,uni7}%
.\newline

In this regard, the Chaplygin gas (CG) model~\cite{refum1,refum2} and its
subsequent generalizations~\cite{refum3,gcg1,gcg2,ngcg} have been proposed as
viable candidates for such a unified description. The CG unifies the two
components through an exotic EoS, exhibiting pressureless matter-like behavior
in the early cosmological era and approaching a cosmological constant during
the late phase of cosmic evolution. However, the standard CG model faces
strong constraints from cosmological observations, particularly from CMB
anisotropies \cite{cgdb1,cgdb2}.\newline

To address these challenges, the generalized Chaplygin gas (GCG) model was
proposed (see, e.g., Refs. \cite{refum1,gcg1,gcg2}) having EoS
\begin{equation}
p_{gcg}=-\frac{\mathcal{A}}{\rho_{gcg}^{\alpha}}%
\end{equation}
In the above expression, $p_{gcg}$ represents the pressure exerted by the fluid, while $\rho_{gcg}$ is the energy density of the fluid. The parameter
$\mathcal{A}$ is a positive constant that indicates the relation between the
pressure and the energy density, and $\alpha$ is taken to be a real number. The cosmological constant is recovered when $\mathcal{A}=1, \alpha=-1$, while the standard CG model is obtained in the special case where
parameter $\alpha$ reaches the limit $\alpha\rightarrow{1}$. Similar to the CG scenario, the GCG model exhibits a smooth transition from a matter-dominated
phase at early cosmic times to a cosmological-constant-dominated phase in the
late cosmological era. This model has been extensively studied by several authors and has been analyzed using a wide range of observational data
\cite{gcgobs1,gcgobs2,gcgobs3,gcgobs4,gcgobs5,gcgobs6}. Interestingly, the GCG
scenario can also be viewed as an interacting $\Lambda$CDM framework in which
a cosmological-constant-like DE exchanges energy with cold DM \cite{gcgobs6}.
The latter scenario is equivalent to the so-called varying vacuum theories.
Because the EoS of DE is not precisely determined and current observations
allow a range $-1.2\leq\omega_{X}\leq-0.61$ \cite{refwx}, it is reasonable to
further generalize the model to accommodate a more general DE component. Apart from describing the late-time Universe, fluids with Chaplygin gas (CG)-like equations of state have also been widely employed in models of cosmic inflation. CG-like models provide exact
solutions for the background cosmological equations where the Hubble parameter and all the physical variables can be expressed by using closed-form functions. The intermediate inflationary universe \cite{Barrow:1990td} is described by a CG-like model, where a cosmological constant-like term is
introduced in the EoS. For other generalizations and observational constraints on the inflationary parameters in CG-like models, see
\cite{Barrow:2016qkh}.\newline

In this direction, an extended framework was introduced in Ref. \cite{ngcg},
commonly referred to as the new generalized Chaplygin gas (NGCG) model, which
provides a unified formulation of DE and DM and can also be interpreted as an
interacting XCDM model. Within this description, the energy exchange between
the dark components is governed by a constant EoS parameter $\omega_{X}$. A
comprehensive discussion of this model is presented in Section~\ref{sec2}, and
further cosmological applications can be found in Refs.
\cite{ngcg01,ngcg02,ngcg03,ngcg04,ngcg04a,ngcgaam1,ngcgaam2}. Recent
observational studies have imposed tight constraints on the NGCG model
parameters. For instance, Salahedin \textit{et al.} \cite{ngcg01} performed a
MCMC analysis by combining multiple observational probes such as supernova
type Ia (SNIa), cosmic microwave background (CMB), baryon acoustic
oscillations (BAO), Big Bang nucleosynthesis (BBN), and Hubble parameter
measurements. Their findings suggest that the NGCG model could offer a viable
approach to alleviating the $H_{0}$ tension in cosmology. Very recently, Mamon
\textit{et al.}~\cite{ngcgaam1} further analyzed the NGCG model with the help
of statefinder and $Om$ diagnostic tools to distinguish it from the standard
$\Lambda$CDM framework and other DE scenarios. In a subsequent
work~\cite{ngcgaam2}, they examined the scalar field correspondence of the
NGCG model and evaluated the associated slow-roll parameters and the resulting
spectral indices. In addition, they analyzed the perturbative behavior of the
model by investigating the evolution of the linear matter density contrast
that governs the growth of matter perturbations. They further discussed the
use of observational growth rate measurements to place constraints on the
model parameters. Finally, their results were compared with the predictions of
the $\Lambda$CDM and GCG frameworks.\newline

Inspired from these considerations, in the present work, we perform a
comprehensive observational analysis of the cosmological scenario arising from
the NGCG framework, by using the most recent background cosmological data.
Specifically, we employ the latest BAO measurements from the Dark Energy
Spectroscopic Instrument (DESI DR2) \cite{DESI:2025zpo,DESI:2025zgx}, combined
with observable Hubble values from cosmic chronometers (CC)
\cite{moresco2020setting}, with the three independent Type Ia supernova (SNIa)
catalogues, the PantheonPlus \cite{Brout:2022vxf}, Union3.0 \cite{union3}, and
DES-Dovekie \cite{DES:2025sig}. By employing three different supernova
samples, we can exclude possible dataset-dependent effects. The recent BAO
data have strengthened the case for a dynamical DE, see for instance Refs.
\cite{Zhang:2025bmk,Nagpal:2025omq,Ormondroyd:2025iaf,You:2025uon,
Wang:2026sqy,Paliathanasis:2025cuc,Scherer:2025esj,
Yang:2025uyv,Santos:2025wiv,Li:2026hwq,Yin:2026gss,Li:2026xaz} and references
therein. In this context, the NGCG scenario is investigated as a unified description of the cosmic fluid, as it provides a simple framework for the dark sector in terms of a single fluid that behaves as pressureless matter at early times and as a DE component at late times. We carry out a Bayesian analysis for the
parameter estimation and a model comparison against the $\Lambda$CDM by using
the Akaike Information Criterion and the Bayesian evidence. The present work
differs from previous observational analyses of the NGCG in two main ways.
We treat the NGCG as a unified dark fluid, allowing the data to constrain the
dark sector directly. Finally, we employ three different SNIa data and we
compare the results, this provides us with information regarding the effects
of the SNIa catalogue on the constraints. The plan of the paper is as
follows.\newline

In Section \ref{sec2} we present a brief discussion of the NGCG model, which
treats DE and DM within a unified framework. Section \ref{sec3} gives an
overview of the observational datasets considered in our study, while
subsection \ref{subsec3.1} presents the statistical methodology together with
the resulting observational constraints. Finally, Section \ref{sec4}
summarizes the main conclusions and provides a brief discussion.


\section{Unifying the Dark Sector with the NGCG}

\label{sec2}
In what follows, we briefly review the NGCG model, which serves as the basis of our cosmological framework and provides a unified-fluid description of the dark sector \cite{ngcg}. The concept of unification of the dark sector has been examined
before in the context of scalar field cosmology. A quintessence unified
scenario was proposed in Ref. \cite{Bento:2004uh}, the reader may also see the discussion in Ref.
\cite{Lukes-Gerakopoulos:2008lvm}. The quintessence unification was recently tested in Ref. \cite{uni4} by using background observations, and it was found
that it can not describe the global dark sector evolution. An alternative
unified scenario was introduced in Ref. \cite{Anagnostopoulos:2019myt} within the
framework of dynamical space-time cosmology. Furthermore, a multiscalar field
unified scenario that can describe DM, DE, and the early inflationary epoch
was proposed in Ref. \cite{Paliathanasis:2023moe}. In this framework, the two
scalar fields interact through both their kinetic and dynamical terms, leading
to a rich cosmological evolution capable of describing the late-time
accelerated expansion as well as the matter-dominated era. During the latter
phase, the nontrivial interaction between the two scalar fields gives rise to the
effective cosmic fluid, providing the mechanism underlying the unified
scenario.\newline

In accordance with the cosmological principle, the large-scale universe is
modeled by a spatially flat, homogeneous and isotropic
Friedmann-Lemaitre-Robertson-Walker (FLRW) spacetime, with the line element given
by
\[
ds^{2}=-dt^{2}+a^{2}\left(  t\right)  \left(  dx^{2}+dy^{2}+dz^{2}\right),
\]
where $a(t)$ is the scale factor of the Universe, and $H(t)=\frac{\dot{a}}{a}$ is the Hubble function which describes the
expansion of the Universe. In this background geometry, the Einstein's
gravitational field equations reduce to a set of ordinary differential
equations known as the Friedmann equations. The field equations relate the
Hubble function and its evolution to the properties of the cosmic fluid. We assume that the latter is described by the NGCG fluid, characterized by an EoS of the form
\cite{ngcg}
\begin{equation}
p_{ngcg}=-\frac{\tilde{\mathcal{A}}(a)}{\rho_{ngcg}^{\alpha}},
\label{eq-rho-gen}%
\end{equation}
where the function $\tilde{\mathcal{A}}(a)$ now evolves with the cosmic scale
factor $a\left(  t\right)  ,$ and $\alpha$ represents the usual Chaplygin gas
parameter.\newline

The NGCG fluid provides a smooth transition between different evolutionary
phases of the cosmos. In early epochs, when $a$ is small, the energy density
evolves as pressure-free matter with $\rho\propto a^{-3}$. On the other hand,
at late times, that is, when $a$ is large, the fluid evolves towards a
DE-dominated regime characterized by $\rho\propto a^{-3(1+w_{de})}$, where
$w_{de}$ represents the EoS of DE. Moreover, the energy density of
the NGCG fluid can be written in a convenient form as \cite{ngcg}
\begin{equation}
\rho_{ngcg}=\left[  Aa^{-3(1+\alpha)(1+w_{de})}+Ba^{-3(1+\alpha)}\right]
^{\frac{1}{(1+\alpha)}}, \label{eqrhongcg1}%
\end{equation}
where the parameters $A,B$ are both positive. It should be emphasized that
Eq.~(\ref{eqrhongcg1}) can be derived by substituting the EoS given in
Eq.~(\ref{eq-rho-gen}) into the continuity equation associated with the NGCG
dark fluid. For consistency, $\tilde{\mathcal{A}}(a)$ must take the form
\begin{equation}
\tilde{\mathcal{A}}(a)=-Aw_{de}\,a^{-3(1+\alpha)(1+w_{de})}.
\end{equation}
Thus, from~(\ref{eqrhongcg1}), the energy density is expressed as
\begin{equation}
\rho_{ngcg}=\rho_{ngcg0}\,a^{-3}\left[  A_{s}a^{-3w_{de}(1+\alpha)}%
+(1-A_{s})\right]  ^{\frac{1}{(1+\alpha)}}.
\end{equation}
Here, $A_{s}=\frac{A}{A+B}$ and $\rho_{ngcg0}=(A+B)^{\frac{1}{1+\alpha}}$
represents the energy density of NGCG at the present epoch.\newline

Since the NGCG framework provides a single description of DM and DE, we can
write $\rho_{ngcg}=\rho_{dm}+\rho_{de}$ and $p_{ngcg}=p_{de}$. The energy
density of DM and that of DE can then be expressed effectively in the form
\cite{ngcg}
\begin{equation}
\rho_{dm}(a)=\rho_{dm0}\,a^{-3}\left[  A_{s}a^{-3w_{de}(1+\alpha)}%
+(1-A_{s})\right]  ^{\frac{1}{(1+\alpha)}-1}, \label{eq-ngcgrdms1}%
\end{equation}

\begin{equation}
\rho_{de}(a)=\rho_{de0}\,a^{-3[w_{de}(1+\alpha)+1]}\left[  A_{s}%
a^{-3w_{de}(1+\alpha)}+(1-A_{s})\right]  ^{\frac{1}{(1+\alpha)}-1},
\label{eq-ngcgrdes1}%
\end{equation}
where $\rho_{dm0}$ and $\rho_{de0}$ correspond to the effective current values
of the energy densities $\rho_{dm}(a)$ and $\rho_{de}(a)$, respectively. It
should be noted that the NGCG model recovers the standard $\Lambda$CDM
cosmology when the parameters are set to $\alpha=0$ and $w_{de}=-1$. In
addition, setting $w_{de}=-1$ recovers the generalized Chaplygin gas model,
while taking $\alpha=0$ corresponds to the $w$CDM model.\newline

From Eqs.~(\ref{eq-ngcgrdms1}) and (\ref{eq-ngcgrdes1}), the two effective
energy densities are related by the formula
\begin{equation}
\frac{\rho_{dm}(a)}{\rho_{de}(a)}=r_{0}\,a^{3(1+\alpha)w_{de}}.
\end{equation}
where $r_{0}=\frac{\rho_{dm0}}{\rho_{de0}}$.\newline

The above relation can be interpreted as describing an energy transfer from one fluid to the other. Indeed, for $\alpha\neq0$, an energy transfer occurs
between the DE and DM components. When $\alpha<0$, we can infer that
there exists an energy flow from DE to DM, whereas for $\alpha>0$, the energy
flow changes direction, and it is transferred from the DM component to the DE.
Consequently, the parameter $\alpha$ quantifies the effective interaction between
the two different components of the NGCG. It should be stressed, however, that
these scalar fields provide only effective representations of the fluid
components, and bear no direct relation to cosmological models with an
interacting sector.\newline

At this point we should remark that the decomposition of the NGCG fluid into
effective dark matter and dark energy components, as presented above, is not
unique. The physical meaningful quantity is the $\rho_{ngcg}$ which enters the
cosmological field equations. Thus, the consideration of the parameter
$\alpha$ as a parameter to quantify the energy transfer between the two
effective components is a feature of the chosen representation for the
decomposition of the fluid. Nevertheless, the results are independent of the
selection of the fluid decomposition.\newline

In the FLRW geometry, the cosmological equations read
\begin{equation}
3H^{2} =\Big[\rho_{dm0}\,a^{-3}+\rho_{de0}\,a^{-3(1+w_{de}\eta)}\Big]\left[
(1-A_{s})+A_{s}\,a^{-3w_{de}\eta}\right]  ^{\frac{1}{\eta}-1}.\nonumber
\end{equation}
where the new parameter is defined as $\eta=1+\alpha$.\newline

We now define the fractional density parameters of the $i$-th component
($i\in\{dm,de\}$) of the Universe at the present epoch as
\begin{equation}
\Omega_{i0}=\frac{\rho_{i0}}{3H_{0}^{2}}.
\end{equation}
where $H_{0}$ is the value for Hubble parameter at the present time.\newline

Substituting these into the Friedmann equation and using $\Omega
_{NGCG0}=\Omega_{dm0}+\Omega_{de0}=1$, the dimensionless Hubble parameter
$E(a)=\frac{H(a)}{H_{0}}$ can be written as
\begin{equation}
E^{2}(a)=a^{-3}\,\Big[\Omega_{dm0}+\Omega_{de0}\,a^{-3w_{de}\eta
}\Big]\,\Big[1-A_{s}\,\big(1-a^{-3w_{de}\eta}\big)\Big]^{\frac{1}{\eta}-1},
\end{equation}
Although commonly adopted in the GCG literature that $A_{s}=\Omega_{de0}$
\cite{ngcg} such that $\Omega_{dm0}=(1-A_{s})$, this is not necessary. In a
unified scenario for the description of the dark sector, it is not required to
provide the definition of either DE or DM.\newline

Furthermore, the above equation simplifies to
\begin{equation}
E^{2}(a)=a^{-3}\,\Big[1-A_{s}\,\big(1-a^{-3w_{de}\eta}\big)\Big]^{\frac
{1}{\eta}}, \label{eqH}%
\end{equation}
which can be equivalently reformulated in terms of the redshift parameter,
$z=\left(  \frac{1}{a}-1\right)  ,$
\begin{equation}
E(z)=(1+z)^{\frac{3}{2}}\left[  1-A_{s}\left(  1-(1+z)^{3w_{de}\eta}\right)
\right]  ^{\frac{1}{2\eta}}. \label{eqHz}%
\end{equation}

We remark that when the parameter $A_{s}=1$, the latter analytic solution
reads~$E(z)=(1+z)^{\frac{3}{2}+3w_{de}}$, which is nothing else than that of
the ideal gas solution. Thus, in the following, we consider $0<A_{s}<1$.

\section{Observational Data Analysis}

\label{sec3}

For the statistical comparison of the unified model with the cosmological
observations, we employ the Bayesian inference framework
Cobaya\footnote{https://cobaya.readthedocs.io/} \cite{cob1,cob2} with the
PolyChord nested sampler \cite{poly1,poly2}, and a custom theory for the
computation of the observables. Furthermore, the GetDist
library\footnote{https://getdist.readthedocs.io/} is applied for the analysis
of the numerical outcomes and plot the confidence space for the posterior
variables~\cite{getd}.\newline

The background observational data considered in this work are:

\begin{itemize}
\item SNIa: We make use of the three Type Ia supernova (SNIa) compilations,
the PantheonPlus (PP) \cite{Brout:2022vxf}, Union3.0 (U3) \cite{union3} and
DES-Dovekie (DESD) \cite{DES:2025sig}. These datasets provide measurements of
the distance modulus $\mu^{obs}$ at different redshifts $z$. The first two
catalogues, that is, the U3 and PP data, share SNIa events within the redshift
range $10^{-3}<z<2.27$. Although these two datasets have 1363 in common SNIa
observational data points, they are constructed from different photometric
reduction pipelines. On the other hand, the DESD catalogue was released
recently after the re-analysis of the five-year Dark Energy Survey supernova
program (DES-SN5YR), yielding 1820 SNIa measurements at lower redshifts
$z<1.13$.

\item CC: We consider the 31 direct measurements of the Hubble parameter
$H\left(  z\right)  $ in the range $0.09\leq z\leq1.965$ as derived by the
cosmic chronometers (CC) \cite{moresco2020setting}.

\item BAO: Finally, we incorporate the DESI DR2 BAO catalogue
\cite{DESI:2025zgx,DESI:2025zpo}, which provides measurements of the
volume-averaged distance ratio, the Hubble distance ratio, and the comoving
angular distance ratio at seven redshifts, each normalized by the sound
horizon at the baryon drag epoch $r_{drag}$ treated here as a free parameter
constrained by the data.
\end{itemize}

For our investigation, we consider three different combinations of the above
datasets, of the form SNIa\&CC\&BAO, where SNIa represent one of the three
catalogues $\left\{  \text{PP,U3,DD}\right\}  $.\newline

The NGCG model has five free parameters, $\left\{  H_{0},r_{drag},A_{s}%
,w_{de},\eta\right\}  $, while the standard $\Lambda$CDM has a
three-dimensional parameter space, consisting of the free variables $\left\{
H_{0},r_{drag},\Omega_{m0}\right\}  $. Therefore, in order to perform a
Bayesian analysis and statistical comparison of the two models, we introduce
the Akaike Information Criterion ($AIC$) and the Bayesian evidence $\ln
Z$.\newline

The PolyChord sampler enables a direct evaluation of the Bayesian evidence,
$\ln Z$. This quantity represents the marginal probability of the data under a
given model and is calculated by integrating the likelihood across the full
parameter space, with each region weighted by its prior probability. The
difference in evidence, $\Delta(\ln Z) = \ln Z_{1} - \ln Z_{2}$, indicates the
degree to which the data statistically favor one model relative to another.
Following Jeffreys' scale \cite{AIC2}, the magnitude of $\Delta(\ln Z)$
determines the strength of model preference. When $\left|  \Delta(\ln Z)
\right|  < 1$, the two models are statistically equivalent in their
description of the data. A value in the interval $1 < \left|  \Delta(\ln Z)
\right|  < 2.5$ corresponds to weak support for the model with the higher
evidence, while $2.5 < \left|  \Delta(\ln Z) \right|  < 5$ signals moderate
preference. If $\left|  \Delta(\ln Z) \right|  > 5$, the evidence in favor of
the model with larger $\ln Z$ is considered strong.\newline

Nevertheless, Akaike's scale uses the $\chi_{\min}^{2}$ and the dimension of
the parametric space. Specifically, from the difference $\Delta\left(
AIC\right)  =AIC_{1}-AIC_{2}$, where $AIC=\chi_{\min}^{2}+2~d.o.f.$, we can
extract information about the statistical preference of the data sets.
Similarly to Jeffreys' scale, if $2<\left\vert \Delta\left(  AIC\right)
\right\vert <4$, the data show weak evidence in favor of the model
with lower $AIC$ parameter, while for $4<\left\vert \Delta\left(  AIC\right)
\right\vert <6$, the support is moderate. Moreover, if the difference is
$\left\vert \Delta AIC\right\vert >6$, there is clear evidence in favor of the
model with lower $AIC$.\newline

For the Bayesian analysis, we considered the priors $H_{0}\in\left[
60,80\right]  $,$~r_{drag}\in\left[  120,170\right]  $, $A_{s}\in\left(
0,1\right)  $, $w_{de}\in\left(  -1.2,-0.5\right)  $, and $\eta\in\left(
0,10\right)  $.%

\begin{table}[tbp] \centering
\caption{Marginalized parameter constraints for the unified NGCG model and statistical comparison with $\Lambda$CDM.}%
\begin{tabular}
[c]{cccc}\hline\hline
\textbf{Dataset} & \textbf{PP\&CC\&BAO} & \textbf{U3\&CC\&BAO} &
\textbf{DD\&CC\&BAO}\\\hline
$\mathbf{H}_{0}$ & $68.0_{-1.7}^{+1.7}$ & $66.9_{-1.8}^{+1.8}$ &
$67.8_{-1.7}^{+1.7}$\\
$\mathbf{r}_{drag}$ & $147.2_{-3.7}^{+3.3}$ & $147.3_{-3.5}^{+3.5}$ &
$147.3_{-3.5}^{+3.5}$\\
$A_{s}$ & $0.819_{-0.063}^{+0.16}$ & $>0.843$ & $>0.822$\\
$\mathbf{w}_{de}$ & $-0.824_{-0.076}^{+0.19}$ & $-0.714_{-0.040}^{+0.17}$ &
\thinspace$-0.768_{-0.051}^{+0.18}$\\
$\eta$ & $1.49_{-0.80}^{+0.30}$ & $1.89_{-1.2}^{+0.55}$ & $1.79_{-1.0}%
^{+0.42}$\\
$\Delta\mathbf{\chi}_{\min}^{2}$ & $-3.32$ & $-8.0$ & $-5.63$\\
$\Delta\mathbf{AIC}$ & $+0.68$ & $-4.0$ & $-1.63$\\
$\Delta\ln\mathbf{Z}$ & $-2.16$ & $-0.13$ & $-0.89$\\\hline\hline
\end{tabular}
\label{tab2}%
\end{table}%

\subsection{Numerical Outcomes}

\label{subsec3.1}

We present below the results derived from the parameter estimation analysis.
We report the median parameter values together with their corresponding 68\%
confidence intervals. In addition, we present the relevant statistical
indicators and employ them to compare the model under consideration with
$\Lambda$CDM, which we adopt as the reference scenario. From the combined
dataset PP\&CC\&BAO, it follows $H_{0}=68.0_{-1.7}^{+1.7}$, $r_{drag}%
=147.2_{-3.7}^{+3.3}$, $A_{s}=0.819_{-0.063}^{+0.16}$, $w_{de}=-0.824_{-0.076}%
^{+0.19}$, and $\eta=1.49_{-0.80}^{+0.30}$. The statistical parameters in
comparison to the $\Lambda$CDM are $\Delta\chi_{\min}^{2}=-3.32,~\Delta
AIC=+0.68$, and $\Delta\ln Z=-2.16$. Thus, while the NGCG model provides a
smaller value for the $\chi_{\min}^{2}$, due to the larger number of the free
parameters, Akaike's scale reveals that the two models are statistically
equivalent, while from the Bayesian evidence, Jeffreys' scale suggests a weak
support of the dataset for the $\Lambda$CDM.\newline

Nevertheless, by introducing the U3 catalogue, that is for the dataset
U3\&CC\&BAO, the analysis of the numerical chains reveals $H_{0}=66.9_{-1.8}%
^{+1.8}$, $r_{drag}=147.3_{-3.5}^{+3.5}$, $A_{s}>0.843$, $w_{de}%
=-0.714_{-0.040}^{+0.17}$ and $\eta=1.89_{-1.2}^{+0.55}$. Moreover, we
calculate $\Delta\chi_{\min}^{2}=-8.0,~\Delta AIC=-4.0$ and $\Delta\ln
Z=-0.13$. Hence, according to the AIC, the dataset provides a weak support in
favor of the NGCG, while Jeffreys' scale indicates that the models are
statistically indistinguishable.\newline

Finally, for the DD\&CC\&BAO combined dataset, we find $H_{0}=67.8_{-1.7}%
^{+1.7}$, $r_{drag}=147.3_{-3.5}^{+3.5}$, $A_{s}>0.822$, $w_{de}=$%
\thinspace$-0.768_{-0.051}^{+0.18}$, and $\eta=1.79_{-1.0}^{+0.42}$, with
$\Delta\chi_{\min}^{2}=-5.63,~\Delta AIC=-1.63$, and $\Delta\ln Z=-0.89$.
Thus, the NGCG provides a smaller value for the $\chi_{\min}^{2}$,
nevertheless from Akaike's and Jeffreys' scales, the two models fit the data in
a similar way.\newline

Overall, while the NGCG model provides smaller values of $\chi_{\min
}^{2}$ than the $\Lambda$ CDM across all the data set
combinations, the NGCG model has two additional parameters, which means that
when the AIC and Bayesian evidence are taken into account, the analysis
reveals that the two models are statistically equivalent.\newline

\begin{figure}[h]
\centering\includegraphics[width=1\textwidth]{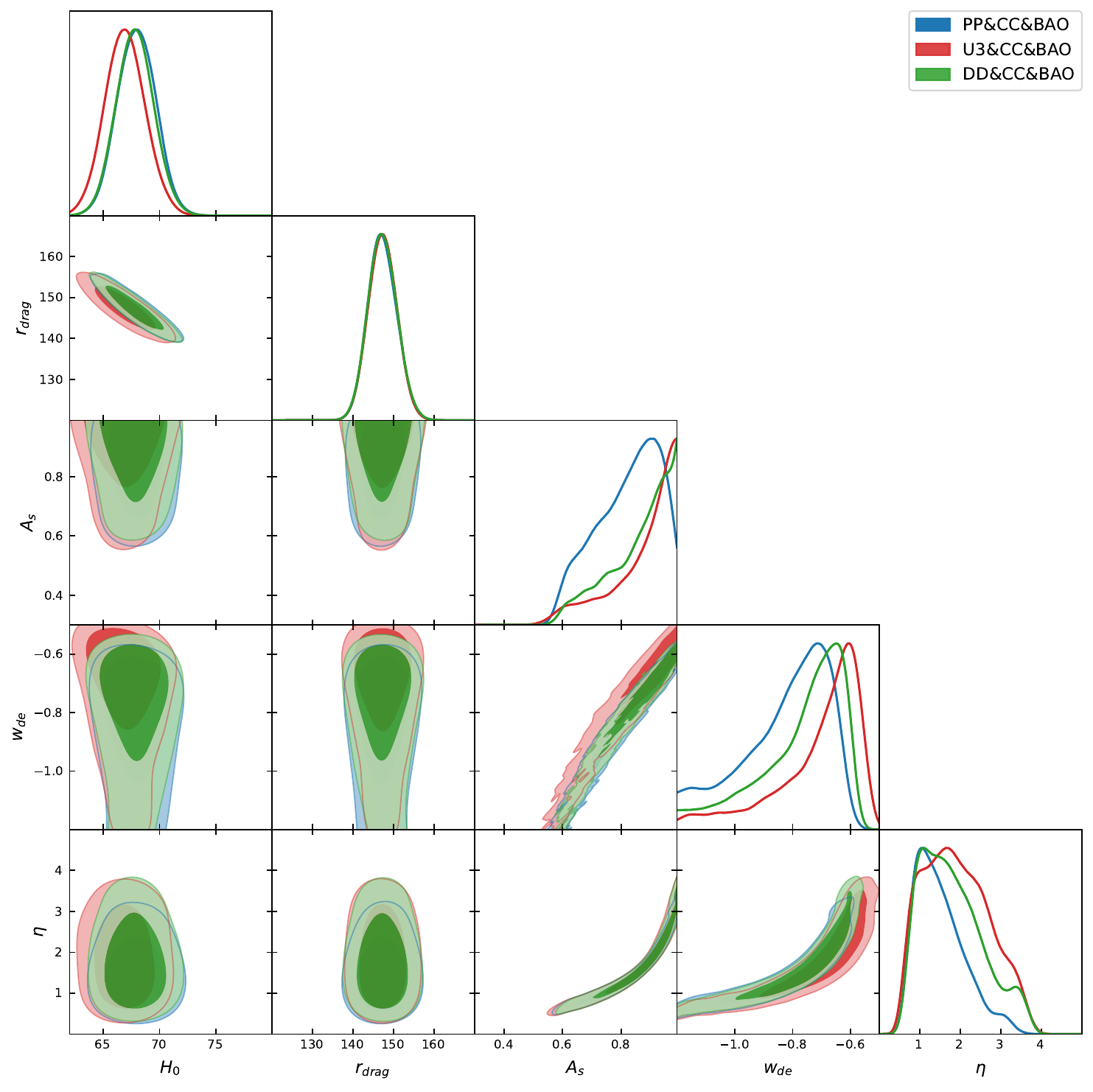}\caption{Marginalized
posterior contours for the parameters \{$H_{0},r_{drag},A_{s},w_{de},\eta$\}
of the NGCG model.}%
\label{fig1}%
\end{figure}

\begin{figure}[h]
\centering\includegraphics[width=1\textwidth]{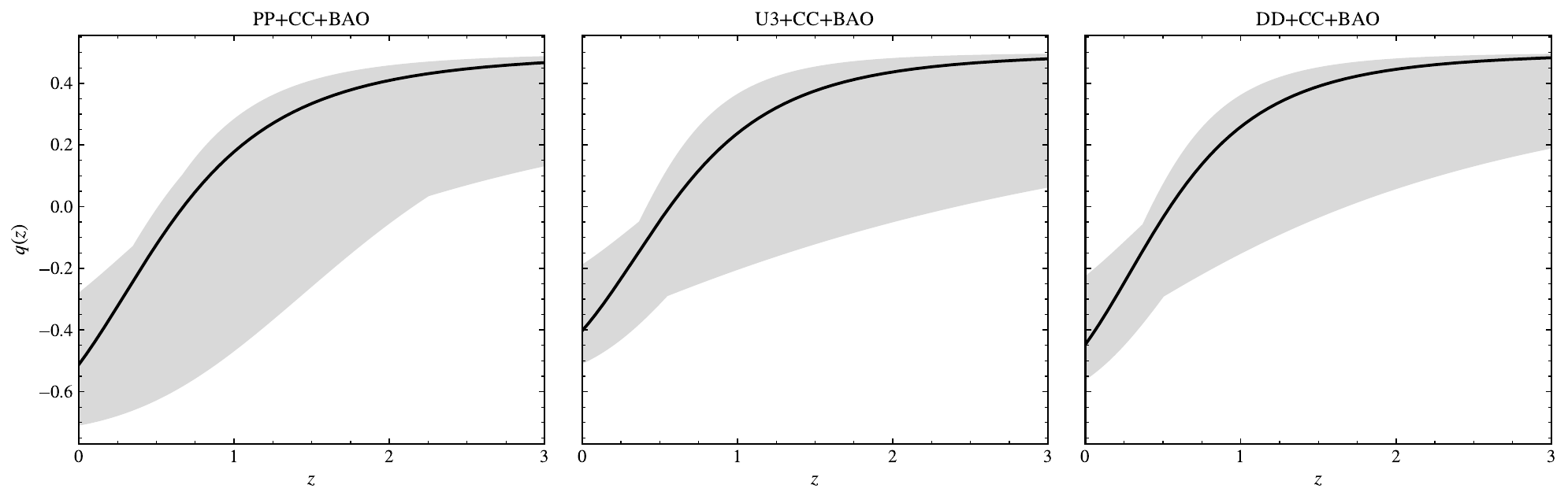}\caption{The
deceleration parameter for the median values and the 68\% confidence interval
for the posterior variables. }%
\label{fig2}%
\end{figure}

\begin{figure}[h]
\centering\includegraphics[width=1\textwidth]{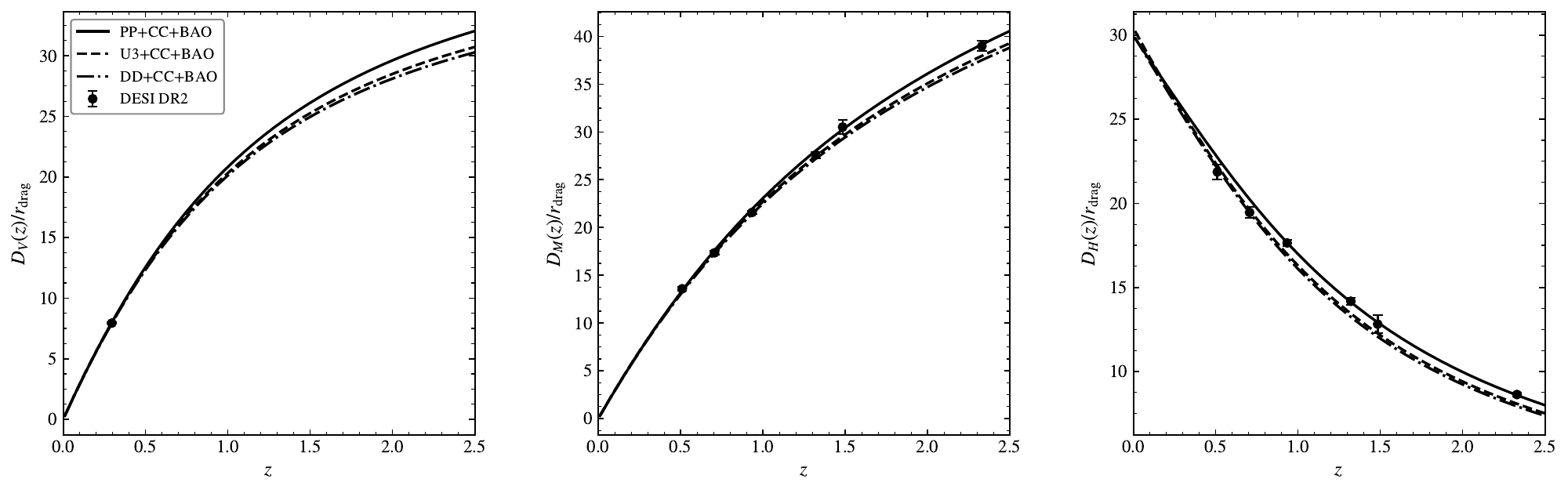}\caption{Evolution of
the theoretical observables for the NGCG with the BAO DESI DR2 data for the
median values of the posterior parameters. The points correspond to the BAO
DESI DR2 measurements are presented with their error bars. }%
\label{fig3}%
\end{figure}

The numerical outcomes from this analysis are summarized in Table \ref{tab2}.
Furthermore, in Fig. \ref{fig1}, we present the contour plots with the
confidence space for the posterior variables. In Fig. \ref{fig2}, we present
the qualitative evolution of the deceleration parameter $q\left(  z\right)  $
for the median values and the 68\% confidence interval for the posterior
variables, with the BAO data of the DESI DR2. Finally, in Fig. \ref{fig3}, we
give the evolution of the theoretical observables with the BAO DESI DR2 data
for the median values of the posterior variables.

\section{Conclusions}

\label{sec4}

We examined the NGCG model as a candidate for describing the dark sector. In
particular, we employed the NGCG as a unified model for DM and DE, where
the effective EoS parameter considered in this work introduces two new free
parameters. The cosmological field equations possess an analytic solution
where the Hubble function can be written in a closed-form expression. We
obtained observational constraints within a Bayesian framework for the
unified model by using late-time background cosmological data. In particular
we employed data combinations of the SNIa compilations, the cosmic
chronometers, and the BAO DESI DR2 measurements.\newline

For the SNIa data, we considered three different catalogues, thus, we
constrained our model with three different combinations of the above described
data sets. The analysis of the numerical chains reveals that the NGCG model
provides systematically lower values for the $\chi_{\min}^{2}$ than the
$\Lambda$CDM. However, because the NGCG model has two additional free
parameters compared to $\Lambda$CDM, when the AIC and the Bayesian evidence
are considered for model comparison to $\Lambda$CDM, the NGCG model is
penalized. Consequently, the NGCG and $\Lambda$CDM models are statistically
indistinguishable according to these criteria.\ Therefore, the background data
applied in this work do not provide any preference for the unified NGCG
scenario over the $\Lambda$CDM.\newline

The NGCG model can also be interpreted as an alternative realization of an
interacting dark sector, with well-defined DM and DE limits. Within this
framework, we find that the median value of the DE EoS parameter lies in the
range $w_{de}\in\left[  -0.824,-0.714\right]  $, suggesting a dynamical DE component rather than a cosmological constant as the late-time
attractor of the unified fluid. Furthermore, the data suggest that the
parameter $\eta$ has a mean value with $\eta>1$ which means
that parameter $\alpha$ is positive, i.e. $\alpha>0$. The
latter can be interpreted as a flow of energy from DM to DE, in the
decomposition considered in this work. This indicates an effective warm dark
matter component, in which the effective dark matter has a non-zero pressure
component. Nevertheless, the limit of $\Lambda$ CDM, i.e $\eta
=1~$ is within the $68\%$ confidence interval. This is
consistent with recent studies, where it was found that pressure or
interaction in the dark matter sector may be favored by current observations
\cite{martens,orl0}.\newline

We emphasize that our analysis is restricted to background data and therefore
does not constrain the NGCG model at the perturbation level. In future work,
we plan to include perturbation observables. This extended analysis will allow
us to investigate whether the model can alleviate current cosmological tensions.

\section*{Data Availability Statement}

All datasets and numerical codes employed in our analysis have been properly
acknowledged, together with the corresponding references.

\begin{acknowledgments}
Part of this study was supported by FONDECYT 1240514. AP thanks Prof. Nikolaos
Dimakis and the Universidad de La Frontera for the hospitality provided when
part of this study was carried out. All the authors are thankful to the anonymous peer-reviewers for their careful
reading of the manuscript and providing many fruitful suggestions for its improvement.
\end{acknowledgments}



\end{document}